# ServiceNet: A P2P Service Network

Ji Liu[1], Hang Zhao[1], Jiyuan Yang[1], Yu Shi[1], Ruichang Liu[1], Dong Yuan[1] and Shiping Chen[1,2]

[1] School of Electrical & Information Engineering, University of Sydney, Australia
[2] Commonwealth Scientific & Industrial Research Organisation (CSIRO), DATA61, Australia
`shiping.chen@data61.csiro.au`

**Abstract.** Given a large number of online services on the Internet, from time to time, people are still struggling to find out the services that they need. On the other hand, when there are considerable research and development on service discovery and service recommendation, most of the related work are centralized and thus suffers inherent shortages of the centralized systems, e.g., adv-driven, lack at trust, transparence and fairness. In this paper, we propose a ServiceNet – a peer-to-peer (P2P) service network for service discovery and service recommendation. ServiceNet is inspired by blockchain technology and aims at providing an open, transparent and self-growth, and self-management service ecosystem. The paper will present the basic idea, an architecture design of the prototype, and an initial implementation and performance evaluation the prototype design.

**Keywords:** Service Discovery, Service Recommendation, P2P, Blockchain.

## 1 Introduction

With the rapid development and evolution of social society and technology innovation, more detailed industry segments have been categorized, such as family wealth management, family health care, personal fitness instructors, etc. Besides, both the service demander and the provider tend to be personalized and individualized, instead of blindly pursuing the traditional unified form. Therefore, a flexible, diverse and fair service docking platform that matches it is needed. This kind of docking platform is most suitable for service-related segments in current world. Furthermore, according to eMarketer Report 2019 for global internet trading market [1], the order number of service-related segments will increase 20% each year, and the order number for 2019 is about 2 billion, and total amount of those orders is around 200 billion dollars. With the huge market capacity and opportunity, there is a rigid demand for a flexible service docking platform.

Traditional service docking platforms are mostly centralized internet platforms [4][5][6]. The owners of these platforms are in charge of all communication and data of both the service requesters and the service providers, so it has asymmetric advantages and rights, which makes it difficult to guarantee: (1) fairness of docking; (2) protection of personal information; (3) and reasonable charges. This leads to mistrust among many parties and affects the efficiency of the entire ecosystem.

Moreover, the personalized recommendation mechanism is not intelligence enough to service current personalized orders for both party of demanders and providers. Hence, a decentralized service docking platform with more intelligence recommendation mechanism feature is urgent for service-related segments.

In this paper, we propose a peer-to-peer (P2P) service network for service discovery and service recommendation, called *ServiceNet*. ServiceNet is inspired by blockchain technology and has the characteristics of decentralization, security, justice, privacy protection, and uses peer-to-peer technology to construct a decentralized service docking network platform to ensure fair docking, privacy protection of data and entities. In addition, ServiceNet leverages some common senses and existing service recommendation techniques to avoid unnecessary intermediate links and services for efficiency, saving bandwidth and minimizing interruption to service providers. Our goal is to create an open, fair, transparent, and intelligent P2P network that truly protects the privacy of the individual, and provides a reliable, efficient, and win-win service docking ecosystem.

The rest of the paper is organized as follows. Section II describes out design principles and the architecture design based on the principles. and Section III talks the implementation of the design as a proof of concept (PoC) prototype. Section IV conducts some performance test and analysis. We present and discuss some related work in Section V. And we conclude in Section VI.

## 2     ServiceNet Architecture Design

We envision a simple motivation example in future to better present our basic idea and design principles for ServiceNet:

*Alice and Bob both use their spare time to provides proof reading services in Sydney and London, respectively. They register their services with ServiceNet as a service provider. Alex and Tom are both self-employed plumbers also in Sydney and London, also registered they services with ServiceNet as a plumber service provider.*

[Use Case 1] *Merry is a PhD student in Sydney. She has completed her PhD thesis and she would like to have her thesis checked before submitting. She sends a thesis proof reading service request to ServiceNet. Since proof reading service can be conducted remotely, both Alice and Bob receive the request, and reply with their quotations. Since Bo's price is a little bit cheaper with a lot of favorable comments, Merry picks up Bob to do the proof reading for her thesis.*

[Use Case 2] *Chen's shower starts leaking. He sends the request to ServiceNet. Since the plumber needs to come Chen's house to fix the shower, ServiceNet only forwards the request to Tom (and maybe a few other plumbers near Chen's home). Through the similar bidding process, Tom secures the deal.*

### 2.1 Design Principles

Based on the above visionary example, we derive our design principles as follows:

- ServiceNet is a P2P network with no centralized servers and business entity, which control ServiceNet.
- It is (almost) free, because of no HR and operational cost.
- It should be smart and fair enough to routine a service request to the less and right service providers, who is likely the best to deliver the service.
- It is able to grow and mange by itself like nature and human society.

### 2.2 ServiceNet Architecture

To follow the above design principle, we design the high-level architecture of ServiceNeT as shown in Fig. 1.

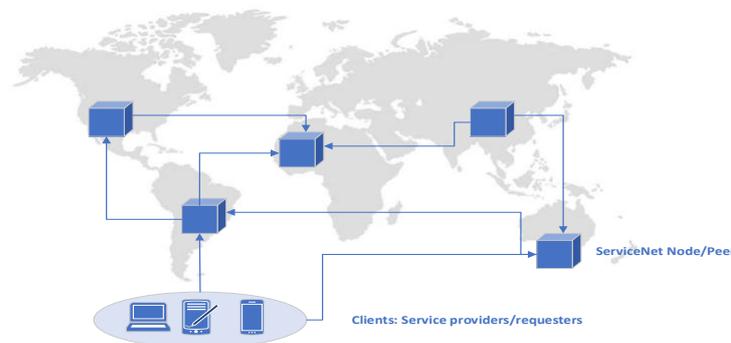

*Fig. 1. Overall architecture of ServiceNet*

As shown in Fig. 1, ServiceNeT consists of a number of nodes (or called peers), which can be deployed onto different continentals and are fully connected each other via the Internet. Each peer has the similar local data and the same functionality, i.e. routine the service requests to the other nodes if required. Clients, either service providers or requesters, can send and receive service-related messages using laptops, iPads, and smart phones via any peer nodes. We provide detailed design for the serviceNet node in the next subsection.

### 2.3 ServiceNet Peer Architecture Design

Each ServiceNet peer node contains at least 3 components to realize the system at our initial design: (a) peer management; (b) Pub/Sub messaging service; and (c) P2P connection service, i.e. the ICE server. While it is feasible to integrate these three components into a single server, we designed them separately for flexible configuration and deployment in future. Fig. 2(a) shows the components of the ServiceNet Peer.

The ICE server is responsible for checking combinations of candidates and establishing a P2P connection between peers. On the other hand, the message transmitting server is a message broker in the Pub/Sub process. It should accept subscription requests from subscribers and route data to them according to the topic when the publishers publish messages. Since the ICE server and the message transmitting server can be complex to be designed and deployed from a fresh start, we should take advantage of matured technologies. Except for the three key attributes, the LTS (Long-Term Support) of technologies chosen should be taken into consideration.

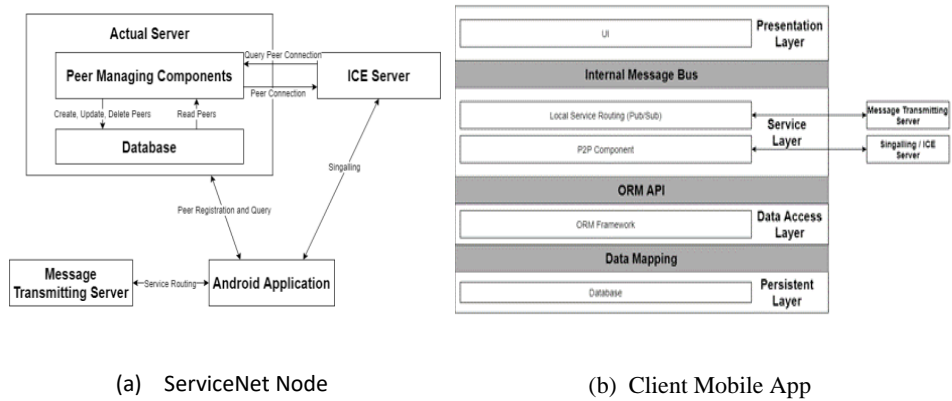

(a)　ServiceNet Node　　　　　　　　(b)　Client Mobile App

*Fig. 2. Architectures of ServiceNet Node and Client Model App*

The actual server is responsible for peer management and WebRTC intermediary signaling support. In terms of peer management, it should in charge of peer registration and keep an instance of peer connection channel. Though the management of peers requires help from persistent data saved in the database, the information stored should be minimized. Thus, the concern of information asymmetry can be mitigated. Besides, it should cooperate with the ICE server to finish the candidates check between peers using the connections it keeps. The initial signaling phase can use the connection channel provided by the actual server to enable metadata exchanging. The overall design of the application layers is following the MVP (Model- View-Presenter) pattern. Components are designed into modules to ensure high scalability and modularity as shown in Fig. 2(b).

**2.3.1 Design of Persistent Layer and Data Access Layer.**

Due to the decentralization nature of the system, the storage load is shifted from the server to peers. All peers are supposed to store their own information and share with others when needed. Besides, synchronization is required for mutual data like chat messages between peers. In terms of the type of the database, we prefer the classical relational database to enable the use of ORM (Object Relational Mapping). An ERD (Entity Relationship Diagram) in Fig. 3 is used to show the database design.

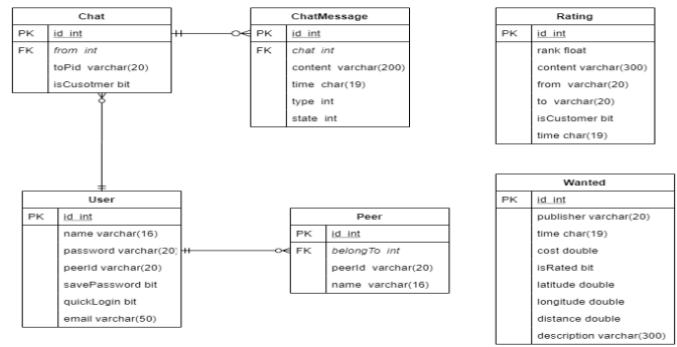

*Fig. 3 Application Database ERD*

PeerId value is used for peer queries in the actual server. This means the querying using such value can be carried on without an actual peer record in the local database. Thus, peerId is not considered as a foreign key in the database. This leads to the disperse of the two tables Rating and Wanted. But a potential relationship does exist among the Peer, the Rating and the Wanted tables.

On the other hand, a data access layer is designed to enable access to the database using OO (Object-Oriented) languages. ORM framework is used to map the relational database into OO objects thus avoid writing of duplicate raw SQL queries. Proper and well-supported ORM framework should be selected to realize the layer. The ORM should also expose pre-defined interfaces to regulate operations to the database.

### 2.3.2 Service Layer.

Service layer is consisting of the P2P connection component and the service routing component. The P2P connection component is mainly responsible for maintaining a connection with the actual server and assisting with the establishment of a P2P connection with others. There is a coupling between the actual server and the ICE server due to the signaling service. Thus, the design of client-side also integrates the functions. Meanwhile, publishing and subscribing messages are handled by the local service routing component. It is the endpoint in the service routing network. It should contain user-defined filters as a local message gateway. Additionally, the two components are supposed to work together to act as the presenter. A response to the user request will be created by querying the local database and exchanging data with the server. The pattern seems similar to the traditional centralized business system, but the difference is that data storage and business logic are injected in the peers. That is to say, the server only transmits messages, the actual work is handled by the peers.

### 2.3.3 Presentation Layer.

The single UI component can be simple to design but complex to implement. Key attributes considered at this stage should be extensibility because we are not focusing on UI design at this stage. The UI should be modulized thus can be easily extended to increase user acceptability in the future. Embedded Ui framework could be considered to fulfil the requirement.

## 3 Implementation of ServiceNet Prototype

### 3.1 ServiceNode Implementation

ICE, NATS Messaging and the actual server can be standalone though we integrated the three servers together. Thus, if such integration becomes a bottleneck, they can be separated into clusters to disperse the load. The former two functional servers are deployed using existing server library NATS Server and Coturn respectively. These libraries expose simple set-up configurations thus can provide high portability and maintainability. Such matured and well-developed applications also provide high availability and reliability.

The actual server is deployed using Node.js. The service layer exposes an API for peer management and signaling assistance. The peer management function involves peer registry, storage, and etc., and is mainly handled in the SocketHandler.js. When a new peer is trying to sign up, the application will generate a UUID (Universally Unique IDentifier) and submit it together with peer's email and name. The UUID is generated using complex data including time and device identifiers etc. by Android provided API. This enables the binding of peer account and the device. Alternation of devices will be detected by the server to protect the account. The server will then generate another two ids called PId (Permanent Id) and TId (Temporary Id) in the service layer. PIds are short, human-readable and unique ids that can identify a peer. They will be generated during peer registry only once. TIds are ids generated by Socket.io to distinguish clients in the session. They are generated each time when the peer comes online. Both ids will be returned to the client via socket. Registered peers can use email or PId to log in anywhere from now on. There is also a map containing peer PId and peer connection instance within the SocketHandler.js for peer management and querying. Peers exchange meta-data in the signaling process with these connection instances. The server can be considered as a yellow page system integrated with the signaling function. Peers can fetch connect information from it then establish a P2P connection with its assistance. In terms of storage, only basic information of peers is stored on the server. The information includes peer PId, nickname, email and UUID. The Constraints.js is created to ease the management of constants like the database connection configuration.

Since we identify availability and scalability as key attributes for the whole system, we postpone error handling to the client instead of in server. If an error emerges during a client request, the error will be reported back to the client for further action.

Scalability is achieved by separating logic. Key logic is being coded into separated.js files and functions are exposed via export keyword. Thus, modules can be changed and rewrite without interfering encapsulation. On the other hand, though the server is written with Node.js, all functions can be easily rewritten to other languages as there is no complex business logic within it and substitution of key libraries can be found in other languages.

### 3.2 Mobile App Implementation

We implemented the client mobile App on Android. Fig. 4 below illustrates the implementation of the Android Mobile App.

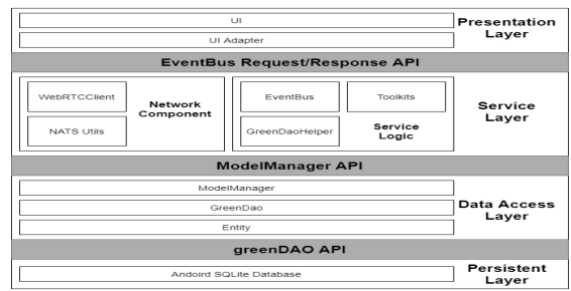

*Fig. 4. Implementation of the Application Layers*

Conforming to the design, the data access layer is designed to assist with accessing the persistent storage of the Android SQLite database by adopting ORM. It is composed of entity classes, greenDAO generated codes and model managers. Entities are POJOs (Plain Old Java Objects) or say Java beans that only contain getters and setters. Each of the class is mapped to a table in the relational database by greenDAO. Additionally, to be able to be transformed into byte streams, they implement the Parcelable interface. By overwriting the abstract methods, instances of the object can be transformed into Parcels. Parcels can be further processed into byte streams thus the instances can be transported through the Internet or Android Bundles. On the other hand, to store database unsupported object, converters are created. Converter are classes overrides specific generic methods defined in the Property Converter interface thus can translate objects into database supported types. For example, Date objects are transformed into strings to enable storage into the database, the StringDateTransformer sample code is shown below Fig. 5 (a)

```
1  class Transformer implements Converter<Date, String
2      @Override
3      public Date convertToProperty(String data) {
4          return StringToDate(databaseValue);
5      }
6
7      @Override
8      public String convertToData(Date property) {
9          return DateToString(entityProperty);
10     }
11 }
```

```
12  ...
13      managers : Map<String, WeakReference<?>>
14      ...
15      chatManager() : ChatManager
16          if (managers.get("chatManager") == null then
17              managers.put("chatManager", new ChatManger
18
19          return managers.get("chatManager")
20  ...
```

(a)                                                  (b)

*Fig. 5. Snapshot of the mobile App implementation*

The greenDAO module is consist of a Database instance, DaoMaster, DaoSession and entity DAOs (Data Access Objects) generated by the greenDAO library. It serves as the ORM API that maps the entities into tables. Connection to Android SQLite database is initiated in the database instance and maintained in the DaoMaster for future access. Transactions and object identity scope are managed in DaoSession. In other words, DaoSession defines Daos and use them to query the desired result from the database. The session can also define whether caches of results are used, that is to say, whether all queries return the reference to the same object. Model managers are encapsulations of entity DAO objects. They expose supported operations to the entities via interfaces. It is the API that the service layer uses to negotiate with the data access layer. All managers should implement a corresponding interface and are managed in the ModelManagerFactory. This fosters modularity and scalability of the module. The factory is a singleton and keeps a map of managers' name and their instances. The instances are created in the form of weak references. This will save RAM and avoid duplicate creation of instances in a short period. The pseudo code is shown in Fig. 5 (b) and the client mobile App UI is shown in Fig. 6(a).

### 3.3 P2P Connection Implementation

The layer mainly consists of two network components and multiple internal service logic components. The WebRTCClient module is a daemon service which will keep running at the backend since the user logged in. It is used to establish and maintain the connection with the actual server and handle P2P connection related operations. It uses EventBus message bus to communicate with other components. In detail, when the user is trying to sign up or log in, the service will be started and try to connect with the server. Then after several message exchanging, it will keep a stable connection with the server. It also supports several query functions which can fetch data from the server. EventBus events are exposed as interfaces for these functions. The sequence diagram in Fig. 6(b) below can show the processes.

Socket.io is used for establishing a connection with the server. We referenced git repository from Naoyuki Kanezawa for Java support [12]. The application can offer high scalability and portability in terms of client-server connection. WebRTCClient is also responsible for establishing a P2P connection with peers. It will use WebRTC's RTCPeerConnection API for signaling and peer instance creation. A peer is abstracted as a Peer object containing the peer's identification, a peer connection instance and a data channel instance. A map with peer's permanent Id and a Peer instance is kept for querying.

NASTS utilities are responsible for connecting with the NATS server, subscribing to a specific channel and receiving published messages from the channel. It serves as the back- bone for the service routing process. Users are required to define filters when subscribing and publishing. The filters will block unqualified messages from users' awareness. Thus, all messages the user attempts to publish or possibly views will be first piped into a filter. This triggers intelligent service matching for individuals. The service logic module contains utility classes for the system. The Eventbus module is designed for application internal message exchange from Greenrobot. Coupling of components can be decreased because direct method calls are transformed to message exchange by EventBus. It adopts the pattern of Pub/Sub to realize internal message exchanging. An object can register and subscribe to a specific kind of event. When another registered object publishes an instance of the event, subscribers can get the instance in order of priority and react to it. There is a concept of "sticky event" which can be received by subscribers even the event is sentin advance. The use of EventBus decreases employment of listeners and overall composition thus increase modularity since all components can work separately with communication only via messages.

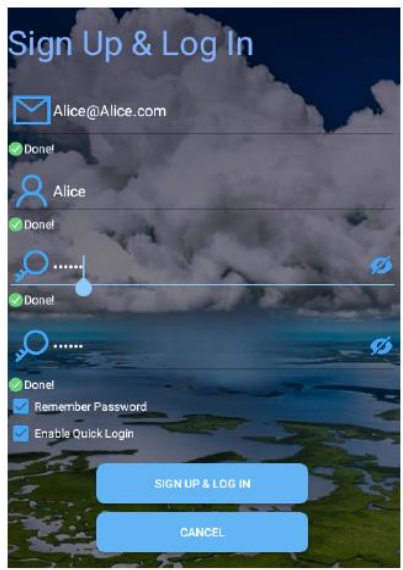
(a) User Mobile App UI

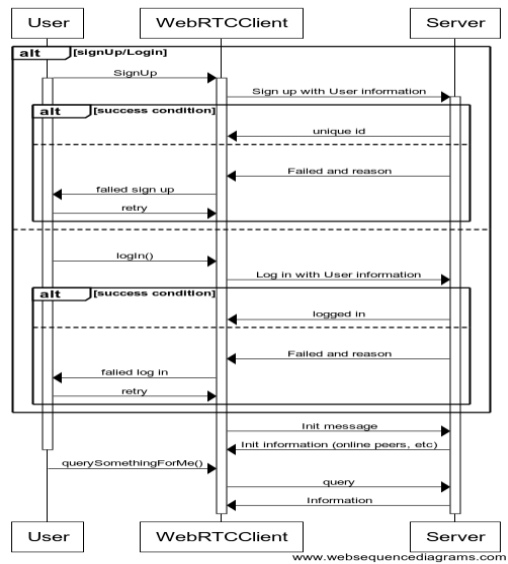
(b) The interaction between the peers

*Fig. 6. The implementation of SericeNet*

GreenDaoHelper helps with initialization of GreenDAO functions. It redefines database update behaviors and provide support for writing and reading encrypted database. The toolkits include multiple utility classes which are mostly static. It integrates commonly used functions like GPS, string transformer, UUID generator, etc.

## 4 Performance Evaluation

We have used scenario analysis, failure analysis, server testing and result analysis, load test, stress test, soak test, etc. to evaluate the performance of our system. Some of the major analysis will introduce below. The performance of the server is tested in terms of response time and through-put. Response time is measured under a variety of circumstances. On the other hand, throughput is used to measure how many requests the server can handle in the given unit of time (normally second).

### 4.1 Test Setting

In terms of test scenario, the server will be tested against the 3 operations identified above. In detail, the registration, login and fetching peer list operations are being tested under different workloads. The workload refers to the client arrival rate in a unit time (second). For example, a test under workload 300 in 10 seconds will result in a total of 3000 (300*10) client's arrival. It should be clarified that there are overlaps between the operations. For instance, in the login operation, the virtual user will first sign up and then login. Such overlap will influence how we treat the statistics. Then the response time and the throughput statistics will be collected and used to draw figures. The response time for login and fetch peers are the total time of accumulation to the previous operations as explained above. That is to say, the response time of login operation is the sum of sign up and the actual login operations. Tests were carried out 5 times under each load and the average result was calculated to gain better accuracy. Moreover, the tests were divided into load, stress and soak test types depending on the period and arrival rate of them. The throughput of the server is recorded during the soak test. The upper bound of server load is identified as 500 which is our ideal maximum client arrival rate in this early stage. The stress test will end at a rate of 1000 clients per second which is enough to test the capability of the server. And the standard arrival rate is set to 300 clients per second.

**Table 1.** Test Environment

| Processor | Intel(R) Core (TM) i7-6700HQ @ 2.6GHz (8 CPUs) |
|---|---|
| RAM | 16384MB |
| Database | MySQL Community Server 5.7.27 |
| Client Emulator | Artillery 1.6.0 |
| Data Generator | Faker.js 4.10.0 |

On the other hand, we encountered issues when selecting test technologies. Though initially, we aimed to simulate the Android environment as real as possible, the capacity of hardware vastly limits the possibility. The cost of starting multiple Android emulators is extremely high thus is not considered. However, even simulating using Socket.io client with Java reveals infeasible after attempts. The high concurrent Socket.io clients will soon fully occupy the CPU. Thus, finally we decided to use Node.js based testing tool Artillery. It can simulate client operations by writing scripts and provides high performance even under high concurrency. Furthermore, for increasing the reality of the simulation, random realistic data is generated by faker.js. The p95 and p99 are selected as attributes in our system. Regarding to the detailed test environment, it is listed together with the computer configuration in Table 1.

### 4.2 Baseline Load Test

The server was tested with 50, 100, 200 and 500 workloads for 10 seconds. Line charts are shown in Fig. 7 (1)-(3). Firstly, the figures are analyzed solely. We can see an obvious boost of response time for all three operations. This is caused by the limitation of the database. The maximum concurrent connection to the local MySQL database is limited to 500. Meanwhile, though there are only 200 clients sending request each second, the server commonly cannot finish all the requests during the one- second interval, Thus the accumulation of requests results in reaching the bound of the database connection. Thus, some of the following requests are required to queue for database access. On the other hand, the increase of response time reaches its peak at 300-500 clients/s load. This is where the server is saturated. This means all incoming requests are immediately queued until another request is finished and a database connection is available.

Then the figures are compared with each other crosswise. There is a notable increase in response time between the registration and the other two operations. This is due to the increased query to the database. Because login needs to firstly sign up and then verify the login information, which is in total 2 queries, thus, can lead to longer response time. The double of queries also leads to a quick boost of the time as shown in the figures. Another interesting point is the increase in response time between login and fetching peers' operations. It increases little when the arrival rate is low. But when the rate is high, the distinction is visible. The storage pattern of peer information leads to the difference. Since the information of peers is attached to the connection instance then stored into a map which is in memory, the querying of the map can be finished in milliseconds when it is small. However, when more clients are online and the volume of the map keeps growing, the query is as efficient. Hence, more key-value pairs can drop into the same bucket in the map thus increase the time finding the desired data.

### 4.3 Stress Test

As 500 clients per second are identified as the ideal upper bound of arrival rate, 600, 700, 800 and 1000 clients per second which exceed the limit were tested. Thus, we can analyze the performance of the server under high stress. The results are shown in Fig. 7 (4)-(6). As shown in the figure, though the stress doubles to 1000clients/s, the response time does not very much. It stays at a similar level as the peek we identified previously. Thus, we believe that the server will not reveal further sudden response time spike under high stress in the current stage. Moreover, the server never fails or produce faults during the stress test such the availability and reliability can be confirmed. One thing worth noting is the gap of p95 and p99 in the peer fetching operation. This is caused by the map issue identified above. Some key-value pairs fall in the same array in the map thus leads to longer response time.

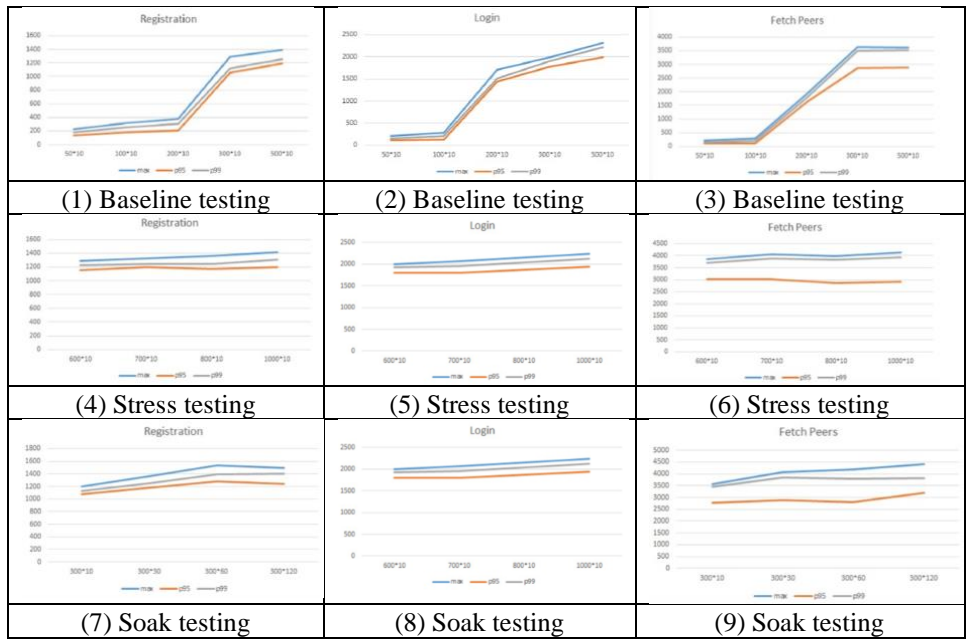

Fig. 7. Put the testing results all together

### 4.4 Soak Test.

The result of soak test is shown in Fig. 7 (7)-(9). This result does not expose much concern as no failure nor a sudden vary of response time occurred. There is a small decrease in time when the time scale is enlarged, this might due to the repetition rate of data generated by faker.js. As time elapses, more identical data can be generated thus result in more rejection when registration. This further leads to immediately fail when the virtual client tries to login or getting peers information. On the other hand, because the three operations have over-laps, throughput is subtracted to gain a clearer image. For example, the final throughput of login is the product of login throughput subtracts registration throughput, it basically remains around 200-270 requests/s which is acceptable to handle the request at the current stage.

In summary, the server does offer high availability and reliability both in low and high workload. It is acceptable that the server can response in 4 seconds even under high stress for all operations for respond time. Moreover, throughput is enough to handle the workload we expected in such an early stage. The current bottleneck is in the database access.

## 5    Related Work

Some researchers have contributed a lot on the field of services discovery. Kozlenkov al et point that service discovery is regarded as one important aspect while developing the service centric systems. They have proposed a framework and built a prototype for helping architecture-driven service discovery. The design phase of development lifecycle can be reduced significantly by this framework, which can offer the systems with functionalities and satisfy properties and constraints [2].

Hu al et have introduced a decentralized service discovery algorithm, named DSDA, to make service discovery algorithm can be suitable for grid environment [3]. E al et believe that the efficient of service discovery is depending on agents' collaboration and the structure among the parties in a distributed system [4]. They have proposed a self-organization mechanism, which can improve the efficient. Yang al et have illustrated that web service is a new generation of web-based application [5]. Meanwhile, they also point that with the increasing of quality and quantity of services, it is an urgent question to provide appropriate services for personalized demand. For solving above question, they have introduced an ontology-based approach of service recommendation with dynamic programming theory, which is tested to be relatively accurate. Li al et also believe that it is very important to recommend personalized web services to users [6]. They think that the temporal influence is an important key factor of Quality of Service, however, the existed papers all ignore it. Thus, they have tried to add the temporal influences as a factor to predict Quality of Service value, and the results have shown that their method outperforms other existed methods.

Several scholars and authors have formerly presented reviews on several use cases of adopting P2P platform in real world. Kellerer et al have built a P2P service platform for mobiles for new ways of service provisioning [7]. The system they built has a relative lower cost and higher data privacy. In the field of social web services, Pantazoglou and Tsalgatidou have built a P2P based platform for publication and discovery in order to solve the problem they found [8]. The platform combines the features of decentralization of P2P and maintenance of Web service descriptions, introduces social networking idea of interactions between service demanders and service providers via collective intelligence emerging. Graffi et al have introduced a life social network platform based on P2P [9]. Liu et al have proposed an improvement of JXTA-Ovarlay with the idea of P2P [10]. Kim and Chung have adopted hybrid P2P network in health field, they use it to mine health-risk factors with PHR similarity [11]. There are many papers about adopting P2P platform into real world, some have been proved to be successful, and only limited applications of applying P2P platform for service-related segments.

Our platform absorbs the advantages of past scholars and removes their disadvantages in p2p respect. At the same time, the characteristics of automatic service discovery are added, which improves the performance of the docking platform a lot.

## 6      Conclusion

This paper explored the idea of building a P2P decentralized service docking network. The decentralized service docking platform is designed and implemented by integrating several techniques including publish/subscribe system (NATS), peer-to-peer communication (WebRTC) to support users with flexibility, diversity, and equivalent fairness. This application is implemented on the Android mobile platform utilizing Android Studio for code development. Initial performance tests are also conducted, and the testing results are provided and discussed.

Although there is an initial proof of concept of ServiceNet idea, we foresee this is an interesting topic for research community and industry to further explore its research issues and potential applications along this direction. For example, one immediate future work for us is to provide interlining service recommendation/routing capability using machine learning technology.